# Two-Dimensional Transport Induced Linear Magneto-Resistance in Topological Insulator $Bi_2Se_3$ Nanoribbons

*Hao Tang†, Dong Liang†, Richard L.J. Qiu and Xuan P.A. Gao\**

Department of Physics, Case Western Reserve University, Cleveland, Ohio 44106

†These authors contributed equally to this work.

\*To whom correspondence should be addressed. Email: xuan.gao@case.edu

ABSTRACT We report the study of a novel linear magneto-resistance (MR) under perpendicular magnetic fields in $Bi_2Se_3$ nanoribbons. Through angular dependence magneto-transport experiments, we show that this linear MR is purely due to two-dimensional (2D) transport, in agreement with the recently discovered linear MR from 2D topological surface state in bulk $Bi_2Te_3$, and the linear MR of other gapless semiconductors and graphene. We further show that the linear MR of $Bi_2Se_3$ nanoribbons persists to room temperature, underscoring the potential of exploiting topological insulator nanomaterials for room temperature magneto-electronic applications.

Topological insulators (TI's) are quantum materials with conducting gapless surface state on the surface or edge of insulating bulk,[1-4] holding great promises in the fundamental study of topological ordering in condensed matter systems and applications in spintronic devices for the spin polarized surface state. The spin polarization and suppressed back scattering render 2D topological surface state an attractive platform for high mobility charge and spin transport devices. Recently, $Bi_2Se_3$ and related materials have been proposed[5] and confirmed[6-8] as three-dimensional (3D) TI's with a single Dirac cone for the surface state. Among these materials, $Bi_2Se_3$, which is a pure compound rather than an alloy like $Bi_xSb_{1-x}$,[9] owns a larger bulk band gap (0.3 eV), and is thought to be promising for room temperature applications. Although the existence of topological surface state in $Bi_2Se_3$ has been established by surface sensitive techniques such as the angle-resolved photoemission spectroscopy,[6,7] extracting the transport properties of 2D surface state in 3D TI's has been plagued by the more dominating conductivity from bulk carriers.[10-18] With extremely high surface-to-volume ratio and thus larger surface contribution in transport, nanostructures of TI's are useful to distinguish 2D surface transport from 3D bulk transport in the heated study of topological surface state. Indeed, Aharonov-Bohm (AB) oscillations were discovered in $Bi_2Se_3$ nanoribbons in the parallel magnetic field induced MR, proving the existence of a coherent surface conducting channel.[10] In this study, we explore the magneto-transport phenomena in $Bi_2Se_3$ nanoribbons in magnetic field perpendicular to the surface of nanoribbons and uncover a novel linear MR effect which is only sensitive to the perpendicular component of the magnetic field (*B*) and absent in parallel field. This 2D magneto-transport induced linear MR is weakly temperature dependent and survives at room temperature, suggesting the possibility of using 2D topological surface transport in room temperature magneto-electronic applications.

## RESULTS AND DISCUSSION

Pure $Bi_2Se_3$ nanoribbons are synthesized in a horizontal tube furnace *via* the vapor-liquid-solid mechanism with gold particles as catalysts, similar to literature.[10,11] Typical $Bi_2Se_3$ nanoribbons have thickness ranging from 50-400 nm and widths ranging from 200 nm to several μms, as shown in the scanning electron microscope (SEM) image in Figure.1a. Transmission electron microscope (TEM) image demonstrates that nanoribbons have smooth side walls and flat surfaces, as shown in Figure.1b for a 200 nm wide nanoribbon. Energy dispersive X-ray spectroscopy (EDX) analyses reveal uniform chemical composition with a Bi/Se atomic ratio about 2:3, indicating the stoichiometric $Bi_2Se_3$. High-resolution TEM imaging and 2D Fourier transformed electron diffraction measurements in Figure.1c and d demonstrate that the samples are single-crystalline rhombohedral phase and grow along the $[11\bar{2}0]$ direction. The upper and lower surfaces are the (0001) planes. The as-grown samples are suspended in ethanol by sonication and dispersed on a heavily doped Si substrate with 300 nm $SiO_2$ on its surface. Photolithography is used to pattern four electrodes contacting single nanoribbon, as shown in the SEM picture in Figure.2a inset. The electrodes consist of 150 nm Pd with a 5 nm Ti adhesion layer formed *via* e-beam evaporation and lift-off. Ohmic contacts are obtained without annealing. The transport measurements are performed in a Quantum Design PPMS with standard low frequency lock-in technique. Four-terminal resistance of the nanoribbons is obtained by flowing a current *I* (typically 0.1-1 μA) through the two outer contacts and monitoring the voltage drop *V* between the two inner contacts (typical spacing ~2 μm) (Figure.2a inset).

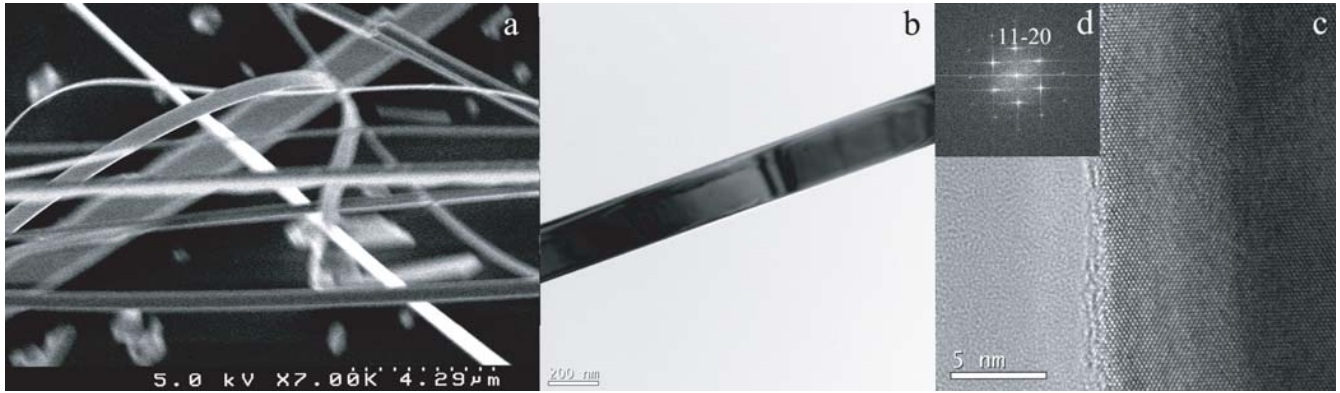

Figure 1. Morphology and crystal structure of $Bi_2Se_3$ nanoribbons. (a) SEM image of as-grown $Bi_2Se_3$ nanowires and ribbons. (b) TEM image showing the shape, flat surfaces and edges of a ribbon with 200 nm width. (c) High-resolution TEM image of the edge of the $Bi_2Se_3$ nanoribbon showing the smooth surface. Scale bar is 5 nm. (d) The Fourier transform electron diffraction pattern indicates the single-crystalline quality of the nanoribbon. The growth direction of the nanoribbon is along $[11\bar{2}0]$.

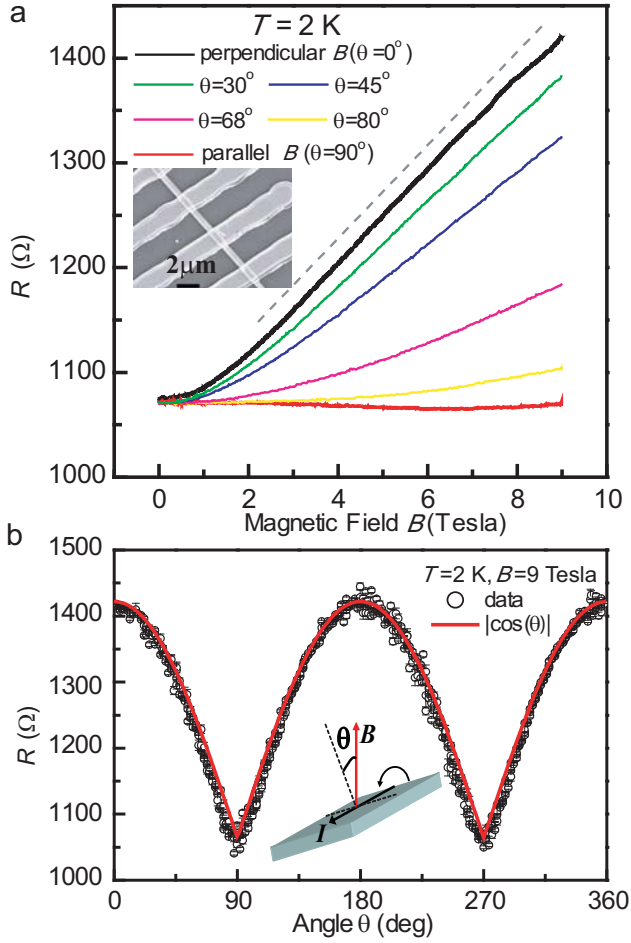

Figure 2. Two-dimensional magneto-transport induced linear magneto-resistance (MR) in $Bi_2Se_3$ nanoribbons. (a) Resistance $R$ vs. magnetic field $B$ of a $Bi_2Se_3$ nanoribbon (sample #1) at 2 K and different tilt angle between sample surface and the magnetic field. The grey dashed line highlights the linear MR above 1 Tesla perpendicular magnetic field. The inset shows SEM image of nanoribbon contacted by four leads in the four-probe transport measurement setup. (b) $R$ vs. the tilt angle $\theta$ in a fixed magnetic field of 9 Tesla and at $T=2$ K. The data are seen to follow the function of $|\cos(\theta)|$ (solid red line). In both (a) and (b), the magnetic field is always perpendicular to the current when the sample is rotated with respect to $B$ (inset).

The resistance $R$ of nanoribbon sample #1 as a function of magnetic field $B$ is shown in Figure. 2a at temperature $T=2$ K. This sample was mounted on a rotating stage such that its surface (a-b plane) could be tilted in magnetic field by an arbitrary angle $\theta$ (Figure.2b inset). Using atomic force microscope (AFM), we measured a sample width of 560 nm and thickness of 100 nm for this nanoribbon. In perpendicular magnetic field ($\theta=0$), the sample exhibited the largest magneto-resistance which is positive and becomes very linear above a characteristic field of 1-2 Tesla. This MR gradually decreased when sample was tilted away from perpendicular configuration and eventually became almost negligible in the parallel magnetic field configuration ($\theta=90º$). It has been known for long time that regular metals exhibit quadratic MR (*i.e.* $\Delta R \propto B^2$) at low fields ($\omega_c \tau < 1$, with $\omega_c$, $\tau$ as the cyclotron frequency and mean scattering time) and this MR would saturate at high field $\omega_c \tau > 1$).[19-21] Therefore, observing a non-saturating linear MR in strong magnetic field is interesting in both the fundamental magneto-transport phenomena and magnetic sensor applications.[19, 22] Very recently, a linear MR was observed in low-doped $Bi_2Te_3$ crystal flakes which showed 2D surface state transport.[18] It is thus encouraging that linear MR is now also observed in topological insulator $Bi_2Se_3$ nanoribbons directly grown from chemical synthesis, in which case the materials can be grown and assembled in large scale for scaled up nanoelectronics applications.[23] The nature of this linear MR in $Bi_2Se_3$ nanoribbons is the focus of this paper.

In the electrical transport study of topological insulator $Bi_2Se_3$ or $Bi_2Te_3$, it is important to distinguish the 3D transport due to bulk carriers from the 2D surface state transport since the Se or Te vacancies may induce high electron concentration and push the Fermi level into the bulk conduction band. One standard way to check if the magneto-transport is 2D or 3D is the angular dependence of the magneto-transport.[13,14,18] For 2D surface state transport, the magneto-transport will only respond to the perpendicular component of the magnetic field $B\cos(\theta)$. In Figure.2b, we show the resistance of sample #1 as a function of rotation angle when the sample was rotated in 9 Tesla magnetic field at 2 K (inset). The measured $R(\theta)$ dependence is seen to have wide peaks around the perpendicular field configuration ($\theta$=0º, 180º) and dips with cusp around the parallel field configuration ($\theta$=90º, 270º). In fact, the experimental data have an excellent agreement with the functional form $|\cos(\theta)|$ (red line in Figure.2b). This suggests that the sample is only responsive to the perpendicular component of *B* at arbitrary tilt angle and thus the linear MR is a 2D magneto-transport effect.

In the strongest perpendicular magnetic fields, $R(B)$ of sample #1 showed weak oscillations on top of the linear MR. These are Shubnikov de-Haas (SdH) oscillations due to the formation of quantized energy levels (Landau Levels) of orbital motions of electrons. We analyze SdH oscillations in another sample (#2) in which these oscillations are more pronounced to extract the carrier density and further demonstrate the 2D nature of Fermi surface. Sample #2 had width of 300 nm and thickness of 25 nm as measured by AFM. Figure.3a shows the *R* vs. *B* from zero to 9 Tesla of sample #2 at various tilt angles. The sheet resistivity $\rho$ is shown on the right side of y-axis, after taking into account the ratio between the length and width of nanoribbon. Similar to sample #1, in perpendicular field, a linear MR was observed above ~1 Tesla. At high fields, the sample exhibited clear SdH oscillations which disappeared quickly once the sampled was tilted away from perpendicular configuration. In parallel field configuration, the MR is negligible, similar to the sample #1. Note that for this experiment, the sample was rotated in the way that the in-plane magnetic field was parallel to the current at $\theta$=90º (Figure.3c inset), in contrast to the rotation scheme for sample #1. However, the magneto-transport effects are qualitatively the same. This leads us to believe that the surface state on side walls of $Bi_2Se_3$ nanoribbons has low mobility and do not contribute much to the observed MR.

To analyze the SdH oscillations, we subtract the linear MR background from the raw $R(B)$ data in Figure.3a and the residual oscillatory part of the MR is shown in Figure.3b for $\theta$=0º, 10º, 22º, 35º (from top to bottom curve). As can be seen, when the sample was tilted, the SdH oscillations move to higher field and eventually our magnetic field (9 T) is not large enough to track the SdH oscillations at $\theta$>35º. By comparing the magnetic field values at the SdH dips we could assign the Landau Level filling factor $\nu$ to be 7 and 8 for the two dips observed at 7.7 and 6.9 Tesla at $\theta=0$. In Figure.3c, we plot the positions of $\nu=7$ and 8 against the tilt angle $\theta$ as sample is tilted away from the perpendicular orientation. The four data points for $\nu=8$ dip are seen to follow the $1/\cos(\theta)$ function nicely. For $\nu=7$, the dip was only observable at $\theta=0º$, 10º, 22º, and moved outside our field range (9 T). However, we infer its position at $\theta=35º$ by multiplying the $\nu=8$ dip position with 8/7. This inferred value together with the three points at lower tilt angle also obeys the $1/\cos(\theta)$ dependence. The agreement between the $1/\cos(\theta)$ function (black lines) and our angle dependent SdH position suggests that the observed SdH oscillations are likely to originate from a 2D Fermi surface.

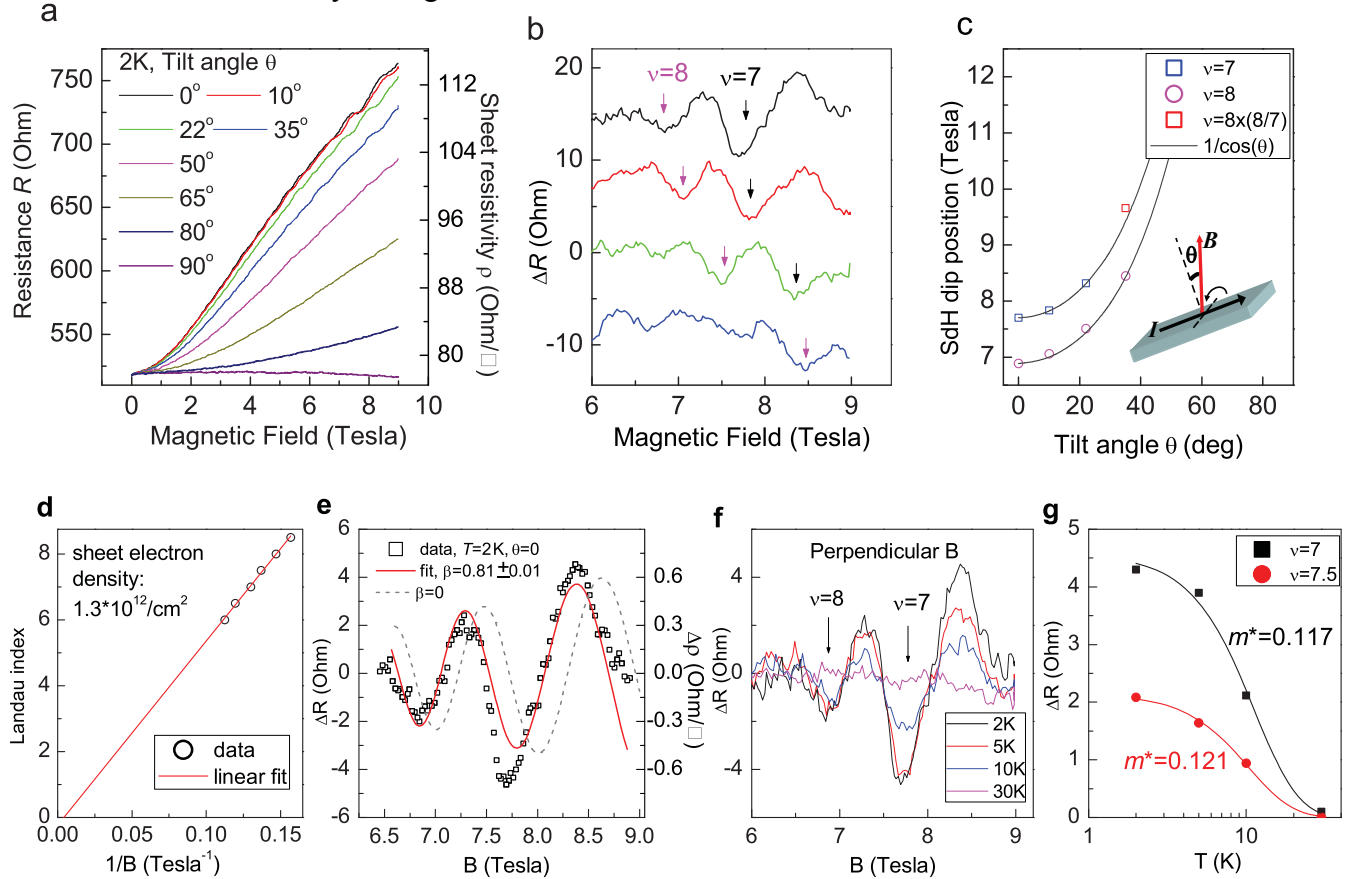

Figure 3. 2D Shubnikov-de Haas oscillations in $Bi_2Se_3$ nanoribbon. (a) MR of sample #2 between 0 and 9 Tesla at $T=2$ K when the rotation angle $\theta$ is increased from 0º to 90º(The rotation scheme is shown as inset of panel (c)). Shubnikov-de Haas (SdH) oscillations are visible on top of the linear MR background for perpendicular field ($\theta=0º$) and disappear rapidly as $\theta$ increases. (b) SdH oscillations at $\theta=0º\rightarrow35º$ after subtracting background MR. The black and pink arrows mark the SdH dips at Landau filling factor $\nu=7$ and 8. (c) Position of the $\nu=7$ or 8 SdH dip plotted against rotation angle. The data are consistent with the $1/\cos(\theta)$ dependence (grey lines) which is expected for 2D SdH oscillations. The inset shows the rotation configuration. (d) Landau index $\nu$ plotted against the inverse of magnetic field for the SdH oscillations in perpendicular field ($\theta=0º$). The slope of linear fit gives a sheet electron density of $n\sim1.3\times10^{12}/cm^2$. (e) Fitting of SdH oscillation at 2 K and $\theta=0$. A non-zero Berry's phase

($\beta$=0.81) is obtained to match the phase of SdH oscillations. (f) SdH oscillations at different $T$. (g) Fitting the amplitudes of SdH dip/peak at $\nu$=7, 7.5 vs. $T$ yields effective electron mass $m^*$=0.117, 0.121± 0.007× $m_e$.

Plotting the Landau index $\nu$ of SdH peak and dip positions vs. the inverse of magnetic field, $1/B$, we obtain a linear dependence in Figure.3d. The slope of the $\nu$ vs. $1/B$ plot gives a 2D sheet carrier density of $1.3\times10^{12}$/cm$^2$. This carrier density corresponds to a Fermi momentum $k_F = 0.41$ nm$^{-1}$ and Fermi energy $E_F = v_F \hbar k_F = 110$ meV. Based on ARPES results,[6,13] the bottom of conduction band is 205 meV above the Dirac point in Bi$_2$Se$_3$. We thus estimate that the position of the 2D surface state Fermi energy is ~95 meV below the bottom of bulk conduction band, consistent with the fact that 3D magneto-transport is negligible in our samples. While both 2D electrons in conventional semiconductors and 2D Dirac electrons have linear $\nu$ vs. $1/B$ dependence, the different Berry's phase would make the intercept zero for regular 2D electrons and non-zero for Dirac electrons with nonzero Berry's phase[24]. The linear fit of $\nu(1/B)$ in Figure.3d produces an intercept of -0.2±0.2. The relatively large uncertainty is due to the limited number of SdH oscillations observed, making it ambiguous to infer the existence of Berry's phase from a simple analysis of $\nu(1/B)$. In such case, it is more useful to directly compare the magneto-resistance data with the known SdH oscillation of 2D electrons and a non-zero Berry's phase would manifest as a phase shift in the oscillations.[25] The general expression for SdH oscillation is[26]

$$\Delta R(B) = A \exp(-\pi/\mu B) \cos\left[2\pi\left(B_F/B + 1/2 + \beta\right)\right] \quad (1)$$

In Eq.1, $B_F$ is the frequency of SdH oscillation in $1/B$ and $\beta \times 2\pi$ is the Berry's phase. The amplitude $A$ has the following temperature dependence

$$A \propto \frac{2\pi^2 k_B T/\hbar\omega_c}{\sinh(2\pi^2 k_B T/\hbar\omega_c)} \quad (2)$$

Using the $B_F$ extracted from the slope of $\nu(1/B)$ dependence, we fit the residual SdH oscillations in $R(B)$ at $T$=2 K to Eq.1 with three fitting parameters $A$, $\beta$, and $\mu$. The fitting yielded a non-zero Berry's phase $\beta$=0.81±0.01 together with $A$=90±50 Ohm and $\mu$ =0.12±0.02 m$^2$/Vs, as shown by the solid red line in Figure.3e. On the other hand, $\beta$=0 with the same $A$ and $\mu$ would predict an oscillatory pattern (dashed grey line in Figure.3e) clearly shifted from experimental data. For 2D Dirac electrons, the Berry's phase of $\pi$ should lead to $\beta$=0.5. However, extracting Berry's phase from SdH of surface state in Bi compound topological insulators is complicated as the application of magnetic field itself may have influence on the value of $\beta$.[18,27] Therefore, a more systematic study on samples with higher mobility in higher magnetic fields (so there are more SdH oscillations) is needed to clarify the Berry's phase effect in Bi$_2$Se$_3$ nanoribbons. Note that our fitted mobility $\mu$=0.12 m$^2$/Vs should be an underestimate since that mobility $\mu$ in Eq.1 is defined with the quantum life time $\tau_q$ which is normally shorter than the transport scattering time $\tau$ in the standard definition of mobility ($\mu = e\tau/m^*$) due to the dominance of small angle scattering contribution in $\tau$.

Figure.3f shows the temperature effect on SdH oscillations. As expected, SdH oscillation amplitude decreases at higher temperature due to the thermal smearing of Landau levels. Fitting the temperature dependence of SdH amplitude to Eq.2 allows us to extract $m^*$, the effective mass of 2D electrons in our $Bi_2Se_3$ nanoribbon. The experimental amplitudes of the SdH peak/dip at $v$ =7, 7.5 vs. temperature are shown in Figure.3g together with fitting curves. Both fitting results yield $m^* \sim 0.12\, m_e$, with $m_e$ as the free electron mass. This value is in quantitative agreement with recent magneto-transport experiment on surface state in bulk $Bi_2Se_3$ crystals with reduced electron densities.[27]

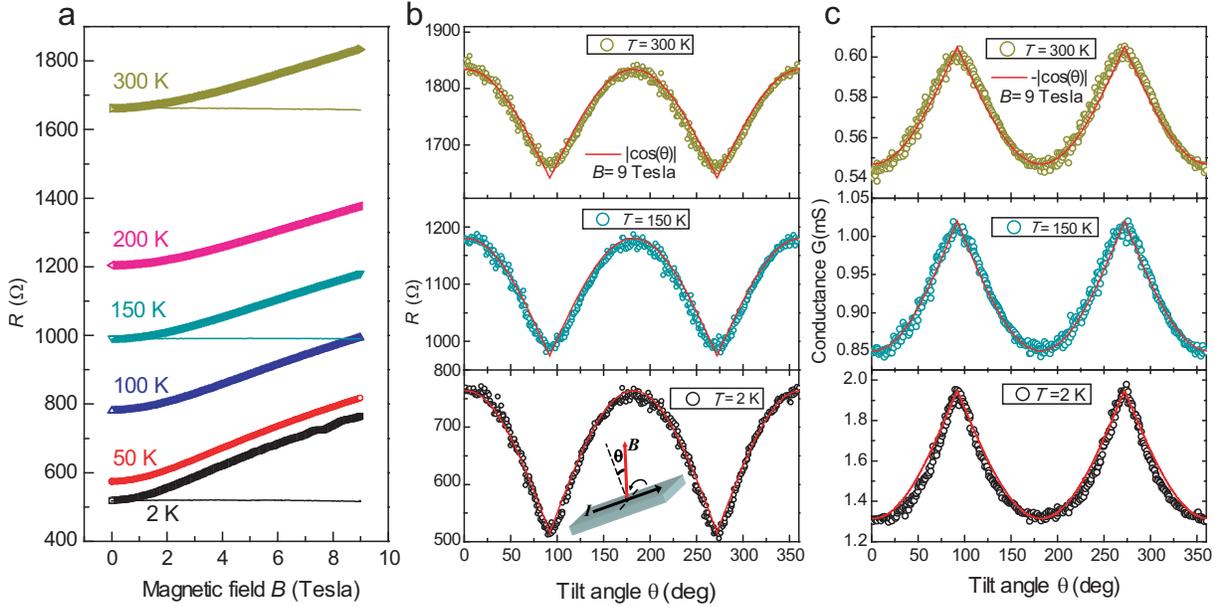

Figure 4. The 2D linear magneto-resistance persisting to room temperature. (a) Resistance vs. magnetic field of sample #2 from $T$=2 K to 300 K. The data in perpendicular field are shown as dots while the data in parallel field are shown as solid lines for $T$=2, 150 and 300 K. (b) $R$ vs. rotation angle at $T$=2, 150 and 300 K. The function $|\cos(\theta)|$ is shown as solid red lines. (c) Conductance of $Bi_2Se_3$ nanoribbon sample #2 vs. rotation angle at $T$=2, 150 and 300 K. Red solid lines represent function $-|\cos(\theta)|$ which is seen to describe the conductance data well for all the temperatures.

Now let us turn back to the main finding of the paper: the linear magneto-resistance induced by 2D magneto-transport. Due to the 2D nature of this linear MR and the associated SdH oscillations, it is likely that this linear MR is intrinsically tied to the 2D Dirac electrons occupying surface state. It would be curious if such surface state magneto-transport can be exploited at high temperature (*e.g.* room temperature) relevant for device applications. We found that raising temperature to room temperature does not have significant influence on the linear MR as shown in Figure.4. In Figure.4a, where the MR of sample #2 is plotted at various $T$ from 2 K to 300 K, one sees that the linear MR at $B_\perp$ >1 T remains unchanged. At the same time, parallel magnetic field induces negligible MR, as shown by the data at $T$=2, 150, 300 K. Angular dependence of MR suggests that the MR is a 2D response over the whole temperature range from 2 K to 300 K, as illustrated by the good agreement between experimental resistance vs. tilt angle data at 9 Tesla and the $|\cos(\theta)|$ function in Figure.4b.

Since both the bulk conduction and surface conduction contribute to the total conductance in $Bi_2Se_3$ crystals, it is more appropriate to discuss conductance $G$ instead of resistance in a quantitative analysis. In Figure.4c, we show $G$ vs. $\theta$ at 2 K, 150 K, and 300 K by inverting resistance data in Figure.4b. It can be seen that the function $-|\cos(\theta)|$ also matches $G(\theta)$ very well. Therefore, the magneto-transport effect we report has 2D nature and is robust from 2 K to room temperature in the magneto-conductance picture. However, due to the increased resistance at high temperatures from phonon scattering, the magneto-conductance at 300 K is nearly ten times smaller than 2 K. This is reasonable given that the thermal energy scale at 300 K already exceeds the maximal cyclotron energy $\hbar\omega_c$ (~100 K) at 9 Tesla, the highest magnetic field in our experiment. If we model the total resistance as $R = 1/(G_b + G_s)$ with $G_b$, $G_s$ as the bulk and surface conductance, a smaller magneto-conductance $\Delta G_s(B)$ for surface state at high temperature is compensated by the smaller value of $G_b$ and still converts into a relatively constant $\Delta R(B)$ in the measurement of total resistance at different $T$. Yet due to the striking robustness of linear MR against raising temperature and the importance of MR in magneto-electronic device applications, here we emphasize the MR instead of magneto-conductance in this paper. We suggest that this linear MR could be exploited in magneto-electronic device operation over wide temperature range and its magnitude is likely to become larger once the nanoribbon crystal quality and mobility improve.[18]

We would like to conclude with a discussion about the possible origins of the linear MR and its relation with other findings in materials with Dirac electronic dispersion spectrum. The well-established Kohler's rule[21] suggests that the MR of a material is a universal function of $\mu B$: $R(B)/R(0) = F(\mu B)$. It is common[21] that at low field when μB<1, $F(\mu B) \approx 1 + (\mu B)^2$, as a result of the Lorentz force deflection of carriers. At high field condition $\mu B$, or $\omega_c\tau > 1$, most materials show saturating MR and a non-saturating and linear MR is unusual. For metals with open Fermi surfaces (*e.g.* Au), the MR could be linear and non-saturating at high fields.[21] This is not the case here for $Bi_2Se_3$. The existence of a linear MR for small bandgap semiconductor could have a quantum[20,28,29] or classical origin.[19,22] To explain the linear MR down to very low fields in silver chalcogenides,[19] Abrikosov first proposed a model based on the quantum-MR[28,29] for systems with gapless linear dispersion spectrum.[28] It is believed that such gapless linear dispersion may apply for silver chalcogenide or other small bandgap semiconductors with strong inhomogeneity.[20, 28] Abrikosov's linear quantum-MR was originally developed at the extreme quantum limit $\hbar\omega_c > E_F$, when all electrons coalesce into the lowest Landau level. Another requirement is that $\omega_c\tau$ or $\mu B > 1$, *i.e.* Landau levels are well-formed. From the above discussions and SdH data, we see that indeed the linear MR occurs at high magnetic field regime where μB>1 (note that our estimate of $\mu$ ~0.12 m$^2$/Vs is an underestimate) and SdH oscillations start to appear. However, there are clearly more than one Landau levels occupied in our sample in the field range where linear MR was observed. This does not exclude the linear quantum MR as the explanation for our data though, since previous experimental study on polycrystal InSb showed that the linear quantum-MR could appear at much lower field with more than one

Landau levels filled. [20] Graphite and epitaxial graphene are also known to exhibit linear MR. [30-33] The linear MR in epitaxial graphene was found to persist to room temperature and interpreted as Abrikosov's linear quantum-MR. [33] However, in order to fully reconcile our experimental data with the linear quantum-MR, further theoretical study is required to understand the fate of linear quantum-MR at high temperatures where $k_B T > \hbar \omega_c$.

Without invoking the linear dispersion spectrum, Parish and Littlewood suggested a classical origin for linear MR in which the MR is a consequence of mobility fluctuations in a strongly inhomogeneous system. [22] For our $Bi_2Se_3$ nanoribbons, the single crystal quality and small length scale of the device rule out the strong physical inhomogeneity of sample surface (AFM imaging also showed only a few quintuple layer step edges over the 1μm length of nanoribbon). However, we can not exclude the possibility of strong electronic inhomogeneity being responsible for the linear MR in our $Bi_2Se_3$ nanoribbon. Electron and hole puddles and charge inhomogeneity have been observed in high quality graphene samples *via* local probe techniques, [34,35] it is quite reasonable that such spatial inhomogeneity of carriers can also exist for gapless surface state on topological insulators.

## CONCLUSION

In summary, we report a linear MR induced by 2D magneto-transport in chemically synthesized nanoribbons of topological insulator $Bi_2Se_3$. When the magnetic field is parallel to the surface of nanoribbon (a-b plane), the MR effect is negligible compared to the MR in perpendicular magnetic fields. Angular dependence of the MR showed an exact |cos($\theta$)| dependence at arbitrary angle, illustrating the 2D origin of linear MR. Angle dependent SdH oscillations also suggest the existence of 2D Fermi surface. These results suggest that the linear MR is likely related to 2D surface state conduction.[36] Furthermore, it is striking that this 2D magneto-transport induced linear MR persists at room temperature, underscoring the potential of exploiting 2D topological surface state for magneto-electronic device application over broad temperature range.

## METHODS

**$Bi_2Se_3$ nanowires and nanoribbons synthesis and characterization.** Pure $Bi_2Se_3$ nanowires and nanoribbons are synthesized in a single zone horizontal tube furnace (Lindberg/Blue M) *via* the vapor-liquid-solid mechanism with gold nanoparticles as catalysts, similar to literature. [10,11] $Bi_2Se_3$ powder is placed and evaporated at the center of a one-inch diameter quartz tube fitted inside the tube furnace to provide the Bi/Se vapor. Silicon (100) substrate functionalized with 10 nm gold nanoparticles (Sigma Aldrich) is placed at the downstream of the Bi/Se vapor flow and used as the wafer supporting the nanowire or nanoribbon growth. Prior to nanoparticle functionalization, silicon substrate was functionalized with 0.1% w/v aqueous poly-L-lysine solution (Ted Pella) to promote the linking of gold nanoparticles to substrate surface. Before nanoribbon growth, the system is pumped to a base pressure of 5 mTorr (limited by the mechanical pump) and flushed with $Ar/H_2$ (10%) gas. $Bi_2Se_3$ powder (99.999% from Alfa Aesar) is evaporated at 680 ºC under the pressure of 200 Torr in 100 sccm flow of $Ar/H_2$ (10%). Typical growth time is 1.5 to 2 hours. After the system cools down, the silicon substrate is covered with a gray coating layer, which consists of $Bi_2Se_3$ nanowires and nanoribbons.

The field-emission gun SEM HITACHI S4500 is used to characterize the morphology of synthesized $Bi_2Se_3$ nanoribbons. The crystal structure and chemical composition of the $Bi_2Se_3$ nanoribbons are characterized by a 300 KV field-emission gun energy-filtering high resolution scanning transmission electron microscope (HRTEM) TECNAI F30, and energy-dispersive X-ray spectroscopy (EDX) equipped in the TEM.

**ACKNOWLEDGEMENT** X.P.A.G. acknowledges P.B. Littlewood, Y. Cui and Y.P. Chen for discussions, ACS Petroleum Research Fund (grant 48800-DNI10) and NSF (grant DMR-0906415) for financial support.

**REFERENCES**

(1) Zhang, S.C. Topological States of Quantum Matter. *Physics* **2008**, 1, 6.

(2) Moore, J.E. The Birth of Topological Insulators. *Nature* **2010**, 464, 194-198.

(3) Hasan, M.Z.; Kane, C.L. Topological Insulators. *Rev. Mod. Phys.* **2010**, 82, 3045-3067.

(4) Qi, X.L.; Zhang, S.C. Topological Insulators and Superconductors. *arXiv*:1008.2026.

(5) Zhang, H.J. ; Liu, C.X. ; Qi, X.L ; Dai, X ; Fang, Z ; Zhang, S.C. Topological Insulators in $Bi_2Se_3$, $Bi_2Te_3$ and $Sb_2Te_3$ with a Single Dirac Cone on the Surface. *Nature Phys.* **2009**, 5, 438-442.

(6) Xia, Y. ; Qian, D.; Hsieh, D.; Wray, L. ; Pal, A. ; Lin, H. ; Bansil, A. ; Grauer, D.; Hor, Y.S. ; Cava, R.J. ; *et al.* Observation of a Large-Gap Topological Insulator Class with a Single Dirac Cone on the Surface. *Nature Phys*. **2009**, 5, 398-402.

(7) Hsieh, D.; Xia, Y.; Qian, D.; Wray, L.; Dil, J.H.; Meier, F.; Osterwalder, J.; Patthey, L.; Checkelsky, J.G.; Ong, N.P.; *et al*. A Tunable Topological Insulator in the Spin Helical Dirac Transport Regime. *Nature* **2009**, 460, 1101-1105.

(8) Chen, Y.L. ; Analytis, J.G., Chu, J.H. ; Liu, Z.K. ; Mo, S.K. ; Qi, X.L. ; Zhang, H.J. ; Lu, D.H. ; Dai, X.; Fang, Z.; *et al*. Experimental Realization of a Three-Dimensional Topological Insulator, $Bi_2Te_3$. *Science* **2009**, 325, 178-181.

(9) Hsieh, D.; Qian, D.; Wray, L; Xia, Y.; Hor, Y.S.; Cava, R.J.; Hasan, M.Z. A Topological Dirac Insulator in a Quantum Spin Hall Phase. *Nature* **2008**, 452, 970-974.

(10) Peng, H.; Lai, K. ; Kong, D. ; Meister, S. ; Chen, Y.L. ; Qi, X.L. ; Zhang, S.C. ; Shen, Z.X. ; Cui, Y. ; Aharonov-Bohm Interference in Topological Insulator Nanoribbons. *Nature Mat.* **2010**, 9, 225-229.

(11) Kong, D.; Randel, J.C. ; Peng, H. ; Cha, J.J. ; Meister, S. ; Lai, K. ; Chen, Y.L. ; Shen, Z.X. ; Manoharan, H.C. ; Cui, Y. Topological Insulator Nanowires and Naonribbons. *Nano Lett.* **2010**, 10, 329-333.

(12) Checkelsky, J.G. ; Hor, Y.S. ; Liu, M.H. ; Qu, D.X. ; Cava, R.J. ; Ong, N.P Quantum Interference in Macroscopic Crystals of Nonmetallic $Bi_2Se_3$. *Phys. Rev. Lett.* **2009**,103, 246601.

(13) Analytis, J.G.; Chu, J.H.; Chen, Y.L.; Corredor, F.; McDonald, R.D.; Shen, Z.X.; Fisher, I.R. Bulk Fermi Surface Coexistence with Dirac Surface State in $Bi_2Se_3$:a Comparison of Photoemission and Shubnikov-De Haas Measurements. *Phys. Rev. B* **2009**, 81, 205407.

(14) Eto, K.; Ren, Z.; Taskin, A.A.; Segawa, K.; Ando, Y. Angular-Dependent Oscillations of the Magnetoresistance in $Bi_2Se_3$ due to the Three-Dimensional Bulk Fermi Surface. *Phys. Rev. B* **2010**, 81, 195309.

(15) Butch, N.P.; Kirshenbaum, K.; Syers, P.; Sushkov, A.B.; Jenkins, G.S.; Drew, H.D.; Paglione, J. Strong Surface Scattering in Ultrahigh Mobility $Bi_2Se_3$ Topological Insulator Crystals. *Phys. Rev. B* **2010**, 81, 241301.

(16) Steinberg, H.; Gardner, D.R.; Lee, Y.S.; Jarillo-Herrero, P. Surface State Transport and Ambipolar Electric Field Effect in $Bi_2Se_3$ Nanodevices. *Nano Letters,* **2010**, 10, 5032-5036.

(17) Checkelsky, J.G.; Hor, Y.S.; Cava, R.J.; Ong, N.P. Surface State Conduction Observed in Voltage-Tuned Crystals of the Topological Insulator $Bi_2Se_3$. *Phys. Rev. Lett.* **2011**, 106, 196801.

(18) Qu, D.X.; Hor, Y.S.; Xiong, J.; Cava, R.J.; Ong, N.P. Quantum Oscillations and Hall Anomaly of Surface States in the Topological Insulator $Bi_2Te_3$. *Science,* **2010**, 329, 821-824.

(19) Xu, R.; Husmann, A.; Rosenbaum, T.F.; Saboungi, M.L.; Enderby, J.E.; Littlewood, P.B. Large Magnetoresistance in Non-Magnetic Silver Chalcogenides. *Nature* **1997**, 390, 57-60.

(20) Hu, J.S. and Rosenbaum, T.F. Classical and Quantum Routes to Linear Magnetoresistance. *Nature Mat.* **2008**, 7, 697-700.

(21) Olsen, J.L. *Electron Transport in Metals*, Interscience, New York, 1962.

(22) Parish, M.M. and Littlewood, P.B. Non-saturating Magnetoresistance in Heavily Disordered Semiconductors. *Nature* **2003**, 426, 162-165.

(23) Fan, Z. ; Ho, J.C. ; Jacobson, Z.A. ; Yerushalmi, R. ; Alley, R.L. ; Razavi, H. ; Javey, A. Wafer-Scale Assembly of Highly Ordered Semiconductor Nanowire Arrays by Contact Printing, *Nano Letters* **2008**, 8, 20-25.

(24) Zhang, Y.B.; Tan, Y.W.; Stormer, H.L.; Kim, P. Experimental Observation of the Quantum Hall Effect and Berry's Phase in Graphene, *Nature* **2005**, 438, 201-204.

(25) Luk'yanchuk, I. A. and Kopelevich, Y. Phase Analysis of Quantum Oscillations in Graphite. *Phys. Rev. Lett.* **2004**, 93, 166402.

(26) Isihara, A. and Smrcka, L. Density and Magnetic Field Dependences of the Conductivity of Two-Dimensional Electron Systems. *J. Phys. C: Solid State Phys.* **1986**, 19, 6777-6789.

(27) Analytis, J.G. ; McDonald, R.D. ; Riggs, S.C. ; Chu, J.H. ; Boebinger, G.S. ; Fisher, I.R. Two-Dimensional Dirac Fermions in a Topological Insulator: Transport in the Quantum Limit. *Nature Physics* **2010**, 6, 960-964.

(28) Abrikosov, A.A. Quantum Magnetoresistance. *Phys. Rev. B* **1998**, 58, 2788-2794.

(29) Abrikosov, A.A. Quantum Linear Magnetoresistance. *Europhys. Lett.* **2000**, 49, 789.

(30) McClure, J.W., Spry, W.J. Linear Magnetoresistance in the quantum limit in graphite, *Phys. Rev.* **1968**, 165, 809.

(31) Zhang,Y.; Small, J.P.; Amori, M. E.S.; Kim, P.; Electric Field Modulation of Galvanomagnetic Properties of Mesoscopic Graphite. *Phys. Rev. Lett.* **2005**, 94, 176803.

(32) Morozov, S.V.; Novoselov, K.S.; Schedin, F.; Jiang, D.; Firsov, A. A.; Geim, A. K. Two-Dimensional Electron and Hole Gases at the Surface of Graphite. *Phys. Rev. B* **2005**, 72, 201401.

(33) Friedman,A.L.; Tedesco, J.L.; Campbell, P.M.; Culbertson, J.C.; Aifer, E.; Perkins, F.K.; Myers-Ward, R.L.; Hite, J.K.; Eddy, C.R.; Jernigan, G.G.; *et al*. Quantum Linear Magnetoresistance in Multilayer Epitaxial Graphene, *Nano Letters* **2010**, 10, 3962-3965.

(34) Martin, J. ; Akerman, N. ; Ulbricht, G. ; Lohmann, T. ; Smet, J.H., Klitzing, K.V. ; Yacoby, A. Observation of Electron-Hole Puddles in Graphene Using a Scanning Single-Electron Transistor. *Nature Phys.* **2008**, 4, 144-148.

(35) Zhang, Y.; Brar, V.W.; Girit, C.; Zettl, A.; Crommie, M.F. Origin of Spatial Charge Inhomogeneity in Graphene. *Nature Phys.* **2009**, 5, 722-726.

(36) Some recent angle dependent SdH and Hall experiments on exfoliated $Bi_2Se_3$ flakes raised the question whether such 2D magneto-transport may be related to the conduction within each individual quintuple layers. (Y.P. Chen, private communication and APS March meeting abstract P35.00012). Such possibility remains to be understood and it is unclear if it is related to our data here.